\documentstyle[12pt,epsfig]{article} 

\newcommand{\beq} 
{\begin{equation}} 
\newcommand{\eeq}{\end{equation}} 
\newcommand{\bea}{\begin{eqnarray}} 
\newcommand{\eea}{\end{eqnarray}}

\def\be{\begin{equation}} 
\def\ee{\end{equation}} 
\def\bea{\begin{eqnarray}} 
\def\eea{\end{eqnarray}} 
 
\begin{document} 
\begin{titlepage}
\title{{\bf Infinite Kinematic Self-Similarity and Perfect Fluid Spacetimes}}
\author{
{Alicia M. Sintes}\thanks{Max-Planck-Institut f\"ur Gravitationsphysik.
  Albert-Einstein-Institut. Am M\"uhlenberg 1, D-14476 Golm. Germany}
 \thanks{Departament de F\'{\i}sica, Universitat de les Illes Balears,  
  E-07071 Palma de Mallorca. SPAIN.}
\and 
{Patricia M. Benoit}\thanks{Department of Mathematics, 
Statistics and Computing Science.
The University of New Brunswick.  Saint John, NB. Canada E2L 4L5}
\and
{Alan A. Coley}\thanks{ Department of Mathematics and Statistics. 
 Dalhousie University. Halifax, NS. Canada B3H 3J5}
}

\date{}
 \maketitle 
\end{titlepage}

\normalsize

\begin{abstract}
Perfect fluid spacetimes admitting a kinematic self-similarity of
infinite type are investigated. In the case of 
plane, spherically or hyperbolically symmetric 
space-times the field
equations reduce to a system of autonomous ordinary differential
equations. The qualitative  properties of solutions of this system of
equations, and in particular  their asymptotic behavior, are studied.
Special cases, including some of the invariant sets and the geodesic
case, are examined in detail and the exact solutions are provided.
The class of  solutions  exhibiting  physical self-similarity  are
found  to play an important role in describing the asymptotic 
behavior of the infinite kinematic self-similar  models.
\end{abstract}


 
\section{Introduction} 
Since the pioneering work of Sedov \cite{Sedov}, the study of self-similar 
systems has played an important role in an extensive range of physical 
phenomena in the classical (Newtonian) theory of continuous media, giving 
rise to many very interesting results with useful experimental and 
astrophysical applications. 
A characteristic of self-similar solutions is that, by a suitable 
transformation of coordinates, the number of independent variables can be 
reduced by one, thus allowing a reduction of the field equations (eg., in some 
cases partial differential equations can be reduced to ordinary DEs). Such 
solutions are of physical relevance since they are often singled out from 
a complicated set of initial conditions; for instance, in an 
explosion in a homogeneous background \cite{Bare72} solutions asymptote to 
self-similar solutions.
 
In the context of general relativity, the concept of self-similarity 
is also largely documented in the literature, beginning with the pioneering 
paper by Cahill and Taub \cite{Cahill71}, and followed by important work by Eardley 
\cite{Eard1,Eard2}.  Spherically symmetric homothetic 
solutions were studied, which proved to be especially useful in the cosmological 
context. More recently Carter and Henriksen \cite{Carter89,Carter91} 
have introduced the concept of kinematic self-similarity, which is a generalization
of the homothetic case.
 
The existence of self-similar solutions of the first kind 
(homothetic solutions) is related to the  
conservation laws and to the invariance of the problem with respect to the  
group of similarity transformations of quantities with independent dimensions.
In this case a certain regularity of the limiting process in passing from the 
original non-self-similar regime to the self-similar regime is implicitly
 assumed.   
However, in general such a passage to this limit need not be regular, whence 
the  
expressions for the self-similar variables are not determined from 
dimensional  
analysis of the problem alone.  Solutions are then called  self-similar 
solutions  
of the second kind.  Kinematic self-similarity is an example of this
more general similarity.  
Characteristic of these solutions is that they  contain  
dimensional constants that are not determined from the conservation  
laws (but can be found by matching the self-similar solutions with the  
non-self-similar solutions whose asymptotes they represent) \cite{Bare72}. 
 
In the study of relativistic dynamics, there is an important  
distinction which must be made.  The existence of a symmetry 
for the geometry (i.e. the metric) does not necessarily imply the existence 
of a symmetry for the matter functions (in particular,
the energy 
density and pressure, when considering a perfect fluid).  For that reason 
it is important to distinguish between the ideas of \lq \lq physical" 
self-similarity 
and \lq \lq geometrical" self-similarity.  The definitions have been given  
elsewhere \cite{Coley96}, and it has been shown that in the case of 
finite kinematic self-similarity the subclass of solutions exhibiting 
physical self-similarity have an important role to play in the examination of 
the full dynamics \cite{Benoit96, Benoit98}.  Similar investigations will 
also be 
important in the study of infinite kinematic self-similarity, and this is
what we shall consider here. 
 
In Benoit and Coley \cite{Benoit96} it was shown that in the case  
of finite kinematic self-similarity
all solutions which exhibit {\it physical} self-similarity asymptote 
(in past and future) to solutions which exhibit a homothety.  In fact, this 
result was extended in Benoit, \cite{Benoit98} to consider all  
solutions with 
finite kinematic self-similarity
 (i.e., not simply those exhibiting {\it physical} self-similarity). 
 
The definitions given previously 
\cite{Carter89,Carter91,Benoit96,Benoit98} 
for the relativistic kinematic self-similarity show the dependence on 
a parameter, commonly denoted by \lq\lq$\alpha$".  In the work of  
Benoit and Coley \cite{Benoit96, Benoit98} the value of this 
parameter is assumed to be an arbitrary {\bf finite} value. 
One special case that was not considered in the previous papers is 
the case in which $\alpha$ takes on an \lq infinite value'.
This case corresponds to the generalization of  rigid 
transformations in general relativity. 
What we attempt in this paper is to study those space-times 
containing a non-null 2-space of constant curvature when they 
admit 
a kinematic self similarity of infinite type, thus complementing 
the previous work.  We shall place special emphasis on models 
which can be interpreted as perfect fluid solutions of Einstein's 
field equations (EFE). In this case the         
 governing system of differential equations reduces to a system of 
autonomous ordinary differential equations and we shall analyze 
the qualitative behavior of these models. 
Exact solutions are obtained in some special cases, particularly
those which are of importance in the asymptotic analysis. 

The paper is organized as follows: Section 2 contains a brief description 
of kinematic self-similarity, deriving the form of the kinematic
self-similar vector field and the self-similar equations for
space-times admitting a three-dimensional  multiply transitive 
group of isometries.
 Section 3 contains the details of the reduction of 
the EFEs 
when there exists a proper kinematic self-similar vector that
commutes with all of the Killing vectors.
  The equations are considered in 
the different cases characterized by the orientation of the fluid flow. 
Section 4 examines the nature of solutions to these equations through 
the use of qualitative methods.  
Section 5 provides the physical asymptotic solutions.
Special cases are then studied in more 
detail in sections 6 and 7. 
 
\section{Kinematic self-similarity and perfect fluids} 
\setcounter{equation}{0} 

A vector field ${\mbox{\boldmath $\xi$}}$ is called a kinematic
self-similar vector (KSS) if it satisfies the conditions \cite{Carter89} 
\be
{\cal L}_{\mbox{\boldmath $\xi$}}u_a=\alpha u_a \quad 
{\rm and}\quad 
{\cal L}_{\mbox{\boldmath $\xi$}}h_{ab}=2\delta h_{ab} \ ,
\label{e2} 
\ee
where $\alpha$ and $\delta$ are constants, ${\cal L}$ stands for the Lie 
derivative operator,
$u^a$ is the four-velocity of the fluid and
$h_{ab}=g_{ab}+u_a u_b $ is the projection tensor which 
represents the projection of the metric into 
the 3-spaces orthogonal to $u^a$.
Evidently, in the case $\alpha=\delta$
it follows that ${\mbox{\boldmath $\xi$}}$ is a homothetic vector (HV)
corresponding to
a self-similarity of the first kind, and if $\alpha=\delta=0$,
${\mbox{\boldmath $\xi$}}$ is a Killing vector (KV).

The similarity transformations are characterized by the scale-independent 
ratio, $\alpha/\delta$, which is 
referred to as the similarity index.  This index is finite except in the 
case of rigid transformations characterized by $\delta=0$.  In this case
the self-similarity is referred
to as \lq infinite' type.
Further information regarding KSS and their properties can be found 
in \cite{Coley96,Sintes98}. 
 
This paper focuses on kinematic self-similar models exhibiting a 
three-dimensional  multiply transitive group of isometries, $G_3$.
Since we consider only perfect fluid models, the $G_3$ 
necessarily acts on space-like orbits $S_2$.
 The solutions 
then correspond to spherical, plane and hyperbolic symmetric      
space-times, and the line element of the metric can be written in 
comoving coordinates as 
\begin{equation} 
ds^2=-e^{2\Phi(t,r)}dt^2+e^{2\Psi(t,r)}dr^2+S^2(t,r)(d\theta^2+ 
\Sigma(\theta,k)^2d\phi^2) \ ,
\label{e4} 
\end{equation} 
where 
\be 
\Sigma(\theta,k)=\left\{ \begin{array}{ll} 
\sin \theta & k=+1 \\ 
\theta & k=0 \\ 
\sinh \theta &k=-1 \ . \end{array} \right. 
\ee 
The four-velocity vector is then given by 
\begin{equation} 
u_a=(-e^{\Phi(t,r)},0,0,0) \ .
\end{equation} 
The Killing vectors (KVs) for the space described by the metric (\ref{e4}) are 
\begin{eqnarray} 
\eta_1&=&sin \phi \partial_\theta + \frac{\Sigma'}{\Sigma}cos 
 \phi \partial_\phi \nonumber \\ 
\eta_2&=&cos \phi \partial_\theta - \frac{\Sigma'}{\Sigma}sin 
 \phi \partial_\phi \nonumber \\ 
\eta_3&=& \partial_\phi \nonumber \ , 
\end{eqnarray} 
where a dash denotes a derivative with respect to $\theta$.
These KVs satisfy the following commutation relations: 
$$                            
[\eta_1,\eta_2]=k\eta_3\ , \quad [\eta_2,\eta_3]=\eta_1 \ , 
\quad [\eta_3,\eta_1]=\eta_2\  . 
$$ 
 
If we assume the existence of a proper KSS, ${\mbox{\boldmath $\xi$}}$, 
then from the Jacobi identities and 
the fact that the Lie bracket of a proper KSS and a KV is  a KV, 
the following algebraic structures are possible
\bea
(I) & & [{\mbox{\boldmath $\xi$}}, \eta_i]=0\ , \quad i=1,2,3\ ,  
\quad k=0,\pm 1 \ , \nonumber \\ 
 & & {\mbox{\boldmath $\xi$}}=\xi^t(t,r)\partial_t+\xi^r(t,r)\partial_r
 \ , \nonumber \\ 
 (II) & & [{\mbox{\boldmath $\xi$}}, \eta_1]=\eta_1\ , \quad
 [{\mbox{\boldmath $\xi$}}, \eta_2]=\eta_2\ ,\quad
 [{\mbox{\boldmath $\xi$}}, \eta_3]=0 \ , \quad k=0 \ ,  \nonumber \\ 
  & & {\mbox{\boldmath $\xi$}}=\xi^t(t,r)\partial_t+\xi^r(t,r)\partial_r
 -\theta \partial_{\theta}
 \ . \nonumber 
\eea
Focusing attention now on equations (\ref{e2}) for the KSS
 and the particular metric (\ref{e4}), it is easy to  
show that the KSS takes the form 
\begin{equation} 
{\mbox{\boldmath $\xi$}}=\xi^t(t)\partial_t+\xi^r(r)\partial_r 
+m\theta \partial_{\theta}\ ,
\end{equation} 
where $m$ is a real number.

In the case in which ${\mbox{\boldmath $\xi$}}$ commutes with all of the KVs,
i.e., $m=0$,
the self-similar equations (\ref{e2}) reduce to 
\bea 
u_{t,t}\xi^t+u_{t,r}\xi^r+u_t{\xi^t}_{,t}& =& \alpha u_t \label{e9} \\ 
h_{rr,t}\xi^t+h_{rr,r}\xi^r+2h_{rr}{\xi^r}_{,r}& =& 2\delta h_{rr} \\ 
h_{\theta\theta,t}\xi^t+h_{\theta\theta,r}\xi^r &=& 2\delta h_{\theta\theta} 
\ , 
 \label{e11} 
\eea 
where a comma indicates partial derivative. 
In this case, the three metric forms (plane, spherical and hyperbolic)
can be studied together. The 
form of the metric functions are similar and the  
EFE's reduce to  single system of ODE's. The second algebraic structure, $(II)$,
is only possible in the plane symmetric case, in which case 
the form of metric
functions and the governing equations are different from those studied here.
 This last case
will be studied elsewhere.

In the infinite case,
$\delta=0$ and ${\mbox{\boldmath $\xi$}}$ can be normalized
so that  the constant $\alpha$ can be set to unity, which we shall do hereafter.
 

\section{Reduction of EFEs } 
\setcounter{equation}{0} 
When attention is restricted  to the case in which the KSS, 
  ${\mbox{\boldmath $\xi$}}$,  commutes with all the KVs,
 (i.e., there exits a proper KSS,  ${\mbox{\boldmath $\xi$}}$,
  orthogonal to all the Killing vectors)
and in which the KSS is of infinite type, three 
 different cases arise.
The three cases are dependent on the orientation of the fluid flow $u$ 
 relative to the KSS; i.e., fluid flow parallel to 
${\mbox{\boldmath $\xi$}}$, fluid flow orthogonal to
 ${\mbox{\boldmath $\xi$}}$, 
 and the most general \lq tilted' case. 
Each case will now be discussed, with the focus and the detailed analysis
made in the general \lq tilted' case.
 
\subsection{Fluid flow parallel to ${\mbox{\boldmath $\xi$}}$} 
 
In this case ${\mbox{\boldmath $\xi$}}$ takes the form  
${\mbox{\boldmath $\xi$}}=\xi^t(t)\partial_t$ 
and without loss of generality we can choose it to be ${\mbox{\boldmath $\xi$}}=t\partial_t$. 
The metric can then be written as 
\begin{equation} 
ds^2=-e^{2\Phi(r)}dt^2+dr^2+S^2(r)(d\theta^2+ 
\Sigma(\theta,k)^2d\phi^2) \ . 
\label{e13} 
\end{equation} 
The EFEs for the perfect fluid become 
\begin{eqnarray} 
\mu S^2 &=& k-S_{,r}^2-2S S_{,rr} \\ 
pS^2 &=& -k+S_{,r}^2+2S S_{,r}\Phi_{,r} \\ 
0 &=& \Phi_{,rr}+\Phi_{,r}^2-\Phi_{,r}\frac{S_{,r}}{S} 
 +\frac{S_{,rr}}{S}-\frac{S_{,r}^2}{S^2}+\frac{k}{S^2} \ , \label{e15}  
 \end{eqnarray} 
where $k=1, 0, -1$ for spherical, plane and hyperbolic symmetry, respectively.
(Note that $\partial_t$ is a KV).  Equation (\ref{e13}) 
 represents a static  space-time.
The function $S_{,r}$ vanishing implies $\mu + p=0$, and therefore 
realistic perfect fluid solutions are excluded. Apart from this case, 
any functions $S$ and $\Phi$ satisfying (\ref{e15}) represent 
kinematic self-similar solutions.

\subsection{Fluid flow orthogonal to ${\mbox{\boldmath $\xi$}}$} 
 
In this case 
${\mbox{\boldmath $\xi$}}$ takes on the form ${\mbox{\boldmath $\xi$}}=\xi^r(r)\partial_r$,
and  without loss of generality we can choose it to be ${\mbox{\boldmath $\xi$}}=\partial_r$.
The metric can then be written as
\begin{equation} 
ds^2=-e^{2r}dt^2+e^{2\Psi(t)}dr^2+S^2(t) 
(d\theta^2+ 
\Sigma(\theta,k)^2d\phi^2) \ . 
\end{equation}     
The field equations for a perfect fluid are 
\begin{eqnarray} 
0&=&S_{,t}  \\ 
0&=&1+ke^{2\Psi}S^{-2}-e^{2\Psi-2r}(\Psi_{,t}^2+\Psi_{,tt}) \\ 
\mu&=&kS^{-2},\quad p=-kS^{-2} \ . 
\end{eqnarray} 
This case is again empty of perfect fluid solutions with $\mu+p \neq 0$. 
 
\subsection{General \lq tilted' case} 
 
The general \lq tilted' case occurs when the four-velocity is neither 
parallel nor orthogonal to the self-similar vector field.  In this case 
one can choose coordinates so that the KSS takes the form: 
\begin{equation} 
{\mbox{\boldmath $\xi$}}=t\partial_t+r\partial_r  \ .
\end{equation} 
In such coordinates, and solving equations (\ref{e9})-(\ref{e11}), it is 
easy to show that the metric can be given by 
\begin{equation} 
ds^2=-e^{2\Phi}dt^2+\frac{e^{2\Psi}}{r^2}dr^2+S^2(d\theta^2+ 
\Sigma(\theta,k)^2d\phi^2) \ ,
\end{equation}    
where  $\Phi, \Psi$ and $S$ are functions depending 
 only on the self-similar 
coordinate 
 \begin{equation} 
\xi=\frac{t}{r} \ .
\end{equation} 
 
The field equations for a perfect fluid source are 
\begin{eqnarray} 
0&=& S''t+S'r-\Phi' S't-\Psi'S't  \label{e20}\\ 
0&=& t^2\Sigma_1(\xi)+\Sigma_2(\xi)\ , \label{e21} 
\end{eqnarray} 
where a dash denotes derivative with respect to $\xi$ and 
\begin{eqnarray} 
\Sigma_1 &\equiv& -(S')^2+\frac{r}{t}S^2\Phi '-S^2\Phi ' \Psi ' +S^2\Phi '' + 
k \frac{r^2}{t^2}e^{2\Psi}+S^2(\Phi ')^2 \\ 
\Sigma_2 &\equiv& e^{2\Psi - 2\Phi}\left[ -\frac{r}{t}SS'-S^2(\Psi ')^2 
+S^2\Phi ' \Psi ' 
 - S^2\Psi ''+(S')^2 \right] \ . 
\end{eqnarray} 
The only possible solutions to equation (\ref{e21}) must necessarily 
satisfy $\Sigma_1=\Sigma_2=0$. 
 
Assuming $\mu+p\neq 0$, we have that $S'$ cannot vanish.  Then defining 
$z\equiv ln(\xi)$ and $\dot{f}\equiv df/dz$, 
 equations (\ref{e20}) and (\ref{e21})
can be rewritten as 
\begin{eqnarray} 
0&=& \frac{\ddot{S}}{\dot{S}}-\dot{\Phi}-\dot{\Psi} \label{e24} \\ 
0&=& -\left( \frac{\dot{S}}{S} \right)^2-\dot{\Phi}\dot{\Psi} 
+k\frac{e^{2\Psi}}{S^2} +\dot{\Phi}^2 \\ 
0&=& -\frac{\dot{S}}{S}+\left(\frac{\dot{S}}{S}\right)^2 
-\dot{\Psi}^2+\dot{\Phi}\dot{\Psi}-\ddot{\Psi}+\dot{\Psi} \ . \label{e26} 
\end{eqnarray} 
Applying now the definitions 
(to be consistent with the notation in \cite{Benoit96})
\begin{equation} 
y \equiv \dot{S}/S \ , \quad u \equiv \dot{\Psi}\ , \quad v \equiv \dot{\Phi}
\ , 
 \quad w \equiv -k e^{2\Psi}S^{-2} \ , \label{defin1} 
\end{equation} 
 equations (\ref{e24})-(\ref{e26}) reduce to
 a 4-dimensional autonomous  system of ODEs 
\begin{eqnarray} 
\dot{y}&=&y(u+v-y) \label{e28}\\ 
\dot{u}&=&-y+y^2-u^2+u+uv \label{e29}\\ 
\dot{v}&=&w+y^2+uv-v^2 \label{e30}\\  
\dot{w}&=&2w(u-y) \ .\label{e31} 
\end{eqnarray} 
The matter quantities are given by 
 \begin{eqnarray} 
\mu&=&e^{-2\Phi}[2yu+y^2]t^{-2}-e^{-2\Psi}[w+2yv+y^2] \label{emu}\\ 
p&=&e^{-2\Phi}[-2yu+2y-y^2]t^{-2}+e^{-2\Psi}[w+2yv+y^2]. \label{epress}
\end{eqnarray} 
Note that the density and pressure can be split as 
$\mu=\mu_1+\mu_2$ and $p=p_1+p_2$ where 
$\mu_1=\hat\mu_1(\xi)t^{-2}$,  $p_1=\hat p_1(\xi)t^{-2}$ and 
 $-p_2=\mu_2=\hat\mu_2(\xi)$. 
Each component of the density and pressure then exhibits 
self-similarity in that  
${\cal L}_{\mbox{\boldmath $\xi$}} \mu_1 = -2\mu_1$, 
${\cal L}_{\mbox{\boldmath $\xi$}} p_1 = -2p_1$ and 
${\cal L}_{\mbox{\boldmath $ 
\xi$}} \mu_2 = {\cal L}_{\mbox{\boldmath $\xi$}} p_2=0$. 
 
We note the following special cases which are evident from equations
(\ref{emu})/(\ref{epress}):
\begin{enumerate}
	\item In the particular case $w+2yv+y^2=0$ 
(i.e., $\mu_2=p_2=0$) the fluid 
is said to be \lq physically' self-similar \cite{Coley96}.
	\item The case $y=0$ is also \lq physically' self-similar;
and since $\mu+p=0$ solutions in this case correspond to 
a cosmological constant solution. 
	\item The case $y=w=0$ gives rise to vacuum solutions. 
	\item Perfect fluid solutions (with $\mu+p \neq 0$) will exhibit 
a barotropic equation of state ($p=p(\mu)$) if and only if 
\begin{equation} 
w+2yv+y^2=c_0e^{2\Psi} \quad {\rm and} \quad 2u+y=c_1 \ ,
\end{equation} 
where $c_0$ and $c_1$ are constants.  
	\item If we are to demand that the solutions satisfy the weak and dominant 
energy conditions (i.e. $-p\leq \mu \leq p$) over the entire manifold 
the following inequalities serve as necessary conditions 
\begin{eqnarray} 
y\geq 0 \ , \nonumber \\ 
y(2u+y-1) \geq 0 \  , \nonumber \\ 
w+2yv+y^2 \leq 0 \ . \label{ec} 
\end{eqnarray} 
We therefore note that by demanding the energy conditions be satisfied 
throughout the evolutions of these models, the possible asymptotic behaviors 
are greatly reduced.
\end{enumerate}
Each of these cases will be important in the analysis of the equations, which follows
in the next sections. 
 

\section{Qualitative analysis} 
\setcounter{equation}{0} 
\label{4}  
 
The system given by equations (\ref{e28}) - (\ref{e31}) is an autonomous 
system of  
first order ODEs.  As such, the asymptotic behavior of the system  
can be determined by studying the qualitative dynamics. 
 
The full system of equations, (\ref{e28}) - (\ref{e31}), describing all 
possible  
solutions, exhibits a number of invariant sets, including the planes 
$$
I_1:\ w=0\ , \quad I_2:\  y=u \ , \quad I_3:\ y=0 \ , 
$$
as well as the surfaces 
$$
I_4:\ w+2yv+y^2=0 \ , \quad I_5: \ w-yv+y^2=0 \ .
$$
To allow for the simplification of the analysis we make the following  
change of variables: 
$$x_1=y, \quad x_2=u - y, \quad x_3=v, \quad x_4=w\ . $$

In these coordinates the equations (\ref{e28}) - (\ref{e31}) become: 
\begin{eqnarray} 
\dot{x_1}&=&x_1(x_2+x_3) \ , \label{5sys1a} \\ 
\dot{x_2}&=&x_2(1+x_3-x_2-3x_1) \ , \\ 
\dot{x_3}&=&x_4+x_1^2+x_1 x_3+x_2 x_3 -x_3^2 \ , \\ 
\dot{x_4}&=&2x_4x_2 \ . \label{5sys1b} 
\end{eqnarray} 

The finite singular points can then be located (note, this system
is not bounded).  They are summarized in 
Table \ref{5tab51}.  There are three distinct  
hyperbolic singular points and two sets of non-isolated singular points, 
each of which have 
zero eigenvalues in the direction tangent to the curve  
and non-zero eigenvalues in all other directions (i.e., they are 
normally hyperbolic).  The finite singular 
points can be classified by the eigenvalues of the Jacobian for the 
vector field.  This classification is given in Table \ref{5tab51}, and 
will be discussed in the sections to follow.   
 
We can also consider the singular points located at infinity.  To do this 
we employ a Poincare transformation 
using the variables: 
\begin{center} 
$X_1=x_1\theta, \quad X_2=x_2\theta, \quad X_3=x_3\theta, 
\quad X_4=x_4\theta$ \ , \\ 
$\theta = (1+x_1^2+x_2^2+x_3^2+x_4^2)^{-1/2} \ .$ 
\end{center} 
In this case the equations (\ref{5sys1a})-(\ref{5sys1b}) become 
\begin{eqnarray} 
X_1'&=& X_1(X_1+X_3-K)-X_1(X_2^2+X_3X_4)\theta, \label{5sysIb}\\ 
X_2'&=& X_2(X_3-X_2-3X_1-K)-X_2(X_2^2+X_3X_4-X_2)\theta, \\ 
X_3'&=& X_1^2+X_3(X_1+X_2-X_3-K)-X_3(X_2^2+X_3X_4-X_4)\theta, \\ 
X_4'&=& X_4(2X_2-K)-X_4(X_2^2+X_3X_4)\theta, \label{5sysIe} 
\end{eqnarray} 
where 
\begin{eqnarray} 
K&=&X_1^2X_2+2X_1^2X_3-3X_1X_2^2+X_1X_3^2-X_2^3 \nonumber \\
  &+&X_2^2X_3+X_2X_3^2-X_3^3+2X_2X_4. 
\end{eqnarray} 
 
The singular points located 
on the invariant boundary $\theta=0$ [the location  
of the infinite singular points for equations (\ref{5sys1a})-(\ref{5sys1b})] 
can be classified by examining the dynamics restricted to this 
invariant surface.   
The location of the singular points 
and their classification are identical to that of the finite case 
\cite{Benoit98}, and are  given in Table \ref{4tab42}. 
 
Returning now to the system (\ref{5sys1a})-(\ref{5sys1b}), we see that 
the invariant hyperplanes $x_4=0$ and $x_2=0$ divide the phase space into  
four additional invariant sets: 
\begin{eqnarray} 
S_1&=&\left\{(x_1,x_2,x_3,x_4)|x_2>0, x_4>0\right\}, \\ 
S_2&=&\left\{(x_1,x_2,x_3,x_4)|x_2>0, x_4<0\right\}, \\ 
S_3&=&\left\{(x_1,x_2,x_3,x_4)|x_2<0, x_4>0\right\}, \\ 
S_4&=&\left\{(x_1,x_2,x_3,x_4)|x_2<0, x_4<0\right\}. 
\end{eqnarray} 
In each of these invariant sets, the function $x_4$ (curvature) 
is monotonic. 
As a result all stable asymptotic behavior is necessarily  
located on one of the invariant sets 
$x_2=0$ or $x_4=0$ (or at $x_4 = \pm \infty$).  Each  
of these  cases will be studied separately.  In all cases we 
note that the classification of the singular points (both finite and infinite) 
can be determined by considering the points listed in 
Tables \ref{5tab51} and \ref{4tab42} 
restricted to the invariant set being considered. 
 
 
\subsection{Subcase: $x_2=0$} 
 
We first consider the hyperplane $x_2=0$.  In this case the system of  
equations (\ref{5sys1a}) - (\ref{5sys1b}) becomes: 
\begin{eqnarray} 
\dot{x_1}&=& x_1 x_3 \ , \label{5sys2a}\\ 
\dot{x_3}&=&w_0+x_1^2+x_1x_3-x_3^2 \ , \label{5sys2b}\\ 
\dot{x_4}&=&0; \quad  x_4=w_0= {\mbox const} \ . 
\end{eqnarray} 
This system is a two-dimensional dynamical system in the variables  
$x_1$ and $x_3$ with parameter $w_0$.  The finite singular points are  
located (where they exist) at: 
\begin{eqnarray} 
L_{1\pm} &=& (0, \pm \sqrt{w_0}), \\ 
L_{2\pm} &=& (\pm \sqrt{-w_0},0). 
\end{eqnarray} 
Each of these points is the intersection of the fixed curves  
($L_1$ and $L_2$ respectively) with the plane under consideration;  
i.e., $x_2=0$ and $x_4=w_0$. 
 
We note here that the value $w_0=0$ is a bifurcation.  We shall first  
consider the dynamics of the solutions when $w_0<0$ and $w_0>0$,  
considering the dynamics at the bifurcation point after. 
We can see from Table \ref{5tab51} that when $w_0<0$ or $w_0>0$, the points 
$L_{1\pm}$ and $L_{2\pm}$ have both positive and negative eigenvalues 
when restricted to this case.  Therefore, each is a two-dimensional saddle point. 
Further, from Table \ref{4tab42} we see that the infinite singular points are 
$B_\pm$, $C_\pm$, $D_\pm$, and $E_\pm$.  Restricted to this invariant 
set we find that $B_+$, $D_-$ and $E_-$ are sources; whereas 
$B_-$, $D_+$ and $E_+$ are sinks. 
 
We now turn our attention to the dynamics of equations 
(\ref{5sys2a})-(\ref{5sys2b}) at the  
bifurcation value of $w_0=0$.  In this case we see that there is only one  
finite singular point, which is located at the origin, $(x_1, x_3) = (0,0)$, 
i.e., at the intersection of the two fixed curves $L_1$ and $L_2$.  
This singular point is non-hyperbolic in nature, and as such its local  
properties can not be determined by examining the eigenvalues of the  
corresponding Jacobian matrix.  In this case, however, there are three  
invariant lines: $x_1=0$, $x_1=-2x_3$ and $x_1=x_3$.  The dynamics on each of 
these lines can be determined as follows: 
\begin{center} 
(i) on $J_1: x_1=0$: $\dot{x_3} = -x_3^2 < 0$. \\ 
(ii) on $J_2: x_1=-2x_3$: $\dot{x_3} = x_3^2 >0$. \\ 
(iii) on $J_3: x_1=x_3$: $\dot{x_3} = x_3^2 >0$. 
\end{center} 
Each of these three invariant lines then divide the 2-dimensional 
phase space into 6 additional invariant regions: 
\begin{center} 
\begin{tabular}{lc} 
$J_4=\{(x_1,x_3)|x_1>0, x_3>x_1 \}$: & $\dot{x_1}>0$ \\ 
$J_5=\{(x_1,x_3)|x_1<0, x_3>-x_1/2 \}$: & $\dot{x_1}<0$ \\ 
$J_6=\{(x_1,x_3)|x_1>0, x_3<-x_1/2 \}$: & $\dot{x_1}>0$ \\ 
$J_7=\{(x_1,x_3)|x_1<0,x_3<x_1 \}$: & $\dot{x_1}<0$ \\ 
$J_8=\{(x_1,x_3)|x_1>0,-x_1/2<x_3<x_1 \}$: & $\dot{x_3}>0$ \\ 
$J_9=\{(x_1,x_3)|x_1<0,x_1<x_3<-x_1/2 \}$: & $\dot{x_3}>0$  
\end{tabular} 
\end{center} 
The result is that the point $(0,0)$ is a saddle. 
The asymptotic analysis is then completed by considering the singular points  
on the infinite boundary.  As the quadratic portion of the vector field is  
unchanged by the differing values of the bifurcation parameter,  
the infinite singular points and the corresponding analysis is identical 
to that when $w_0 \neq 0$. 
 
A bifurcation diagram, including all the phase portraits for each  
range of the parameter $w_0$ is given in Figure \ref{fig:51}.  As can be seen by 
these phase portraits all generic asymptotic behavior (to the past and 
the future) is located on the infinite boundary.  The exact solutions 
for each of these singular points (which are asymptotic states to past or future 
or are intermediate states) will be examined in section 5.

\subsection{Subcase: $x_4=0$ - plane symmetry} 
 
The invariant set $x_4=0$ contains a subset of the asymptotic solutions 
for the system (\ref{5sys1a})-(\ref{5sys1b}).   
As can be seen from equations (\ref{e4}) and (\ref{defin1}), solutions 
which have $w$ identically zero comprise the set of plane symmetric solutions.
 
In this case, the system of ODEs (\ref{5sys1a})-(\ref{5sys1b}) reduces to: 
\begin{eqnarray} 
\dot{x_1}&=&x_1(x_2+x_3) \ , \label{5sys3a}\\ 
\dot{x_2}&=&x_2(1+x_3-3x_1-x_2) \ , \\ 
\dot{x_3}&=&x_1^2+x_3(x_1+x_2-x_3) \ . \label{5sys3b} 
\end{eqnarray} 
 
\noindent 
The co-ordinate planes $x_1=0$ and $ x_2 =0$ are each invariant sets for  
this system, as are the sets $x_1+2x_3=0$ and $x_1=x_3$.   
 
The finite singular points in this case are given by $Q_1$, $Q_2$, $Q_3$ 
and $L_1=L_2$.  The local dynamics of each is determined by considering the 
sign of the eigenvalues of the Jacobian (see Table \ref{5tab51}) restricted to
this set $x_4=0$; i.e, those eigenvalues whose associated eigenvectors 
have the form $(c_1,c_2,c_3,0)$.  The points $Q_1$-$Q_3$ are saddles in  
this three-dimensional set.  The point $L=L_1=L_2$ is non-hyperbolic. 
Center manifold theory \cite{CMT} allows the point to be analyzed.  
The many 
invariant sets which include this point greatly simplify the analysis, 
and it is a straightforward matter to show that in the two dimensions 
which define the coordinate plane $x_2=0$ the point is a saddle and 
in the third direction it is a saddle-node. 
The infinite singular points  
(not including $C_\pm$) are given in Table \ref{4tab42}.
The dynamics on the infinite boundary is  
represented by Figures \ref{fig:45a} and \ref{fig:45b} (see \cite{Benoit98} for details).

Before considering the global dynamics in this three-dimensional system, 
we shall consider the dynamics as restricted to the  invariant planes. 
Each of these planes will divide the phase space further, allowing for a 
simplification in the analysis when considering the entire space. 
Note that the $x_2=0$ invariant set has been completely analyzed in the 
previous section.  The dynamics are represented by the case $w_0=0$ in  
Figure \ref{fig:51}.  Therefore, we need only consider the planes $x_1=0$, 
$x_1+2x_3=0$ and $x_1=x_3$. 
 
\noindent 
{\bf Invariant Set: $x_1=0$} 
 
\noindent 
In the invariant set $x_1=0$, the system (\ref{5sys3a})-(\ref{5sys3b}) reduces
 to: 
\begin{eqnarray} 
\dot{x_2}&=& x_2(1+x_3-x_2) \ , \label{5x1a} \\ 
\dot{x_3}&=&x_3(x_2-x_3)\ , \label{5x1b}  
\end{eqnarray} 
and represents the vacuum solutions in the full 4-dimensional
system.
This system gives rise to dynamics in the $x_2$ - $x_3$ plane.  The finite 
singular points are given by $L_1=L_2=(0,0)$ and $Q_1=(1,0)$.  Local  
analysis shows that the point $(1,0)$ is a saddle point and 
the point $(0,0)$ is non-hyperbolic, saddle-node in nature (determined  
through the use of center manifold theory).  Therefore, 
no stable asymptotic behavior is located in the finite part of the phase space
and all asymptotically stable solutions in this subcase are located on the  
infinite boundary.  The complete phase portrait, as compactified by the Poincare 
transformation, is given in Figure \ref{fig:52}.

\noindent 
{\bf Invariant Set:} $x_1+2x_3=0$ 
 
\noindent 
In the invariant set $x_1+2x_3=0$, the system (\ref{5sys3a})-(\ref{5sys3b}) 
reduces to: 
\begin{eqnarray} 
\dot{x_2}&=& x_2(1+3x_3-x_2), \label{5x13a} \\ 
\dot{x_3}&=&x_3(x_3+x_2). \label{5x13b} 
\end{eqnarray} 
As a result the dynamics is located in a two-dimensional plane.  The finite 
singular points are given by $Q_1=(1,0)$, $Q_3=(1/8, -1/8)$ and $L_1=L_2=(0,0)$.  Local  
analysis determines that the point $(1,0)$ is a saddle, $(1/8, -1/8)$  
a spiraling sink and $(0,0)$ a saddle-node  (determined through the use of  
center manifold theory).  The phase portrait for this case, as compactified by
 the Poincare 
transformation, is given in Figure \ref{fig:53}.  In this case the fluid is 
also physically self-similar.

\noindent 
{\bf Invariant Set:} $x_1=x_3$ 
 
\noindent 
In the invariant set $x_1=x_3$, the system (\ref{5sys3a})-(\ref{5sys3b}) reduces to: 
\begin{eqnarray} 
\dot{x_1}&=& x_1(x_1+x_2), \label{5x3a}\\ 
\dot{x_2}&=&x_2(1-x_2-2x_1). \label{5x3b} 
\end{eqnarray} 
As a result the dynamics are located in a two-dimensional plane.  The finite 
singular points are given by $L_1=L_2=(0,0)$, $Q_1=(0,1)$ and $Q_2=(1, -1)$.  
Local  
analysis determines that the points $(1,0)$ and $(1,-1)$ are saddles 
whereas $(0,0)$ is a saddle-node (determined through the use of  
center manifold theory).   
The phase portrait for this case, as compactified by the Poincare 
transformation, is given in Figure \ref{fig:54}.

\noindent 
{\bf Global dynamics} 
 
\noindent 
The global dynamics can be determined through an investigation of 
the direction fields making use of the monotonicity principle \cite{LeBlanc}. 
To simplify the global analysis, consider the full three-dimensional 
phase space divided into 16 invariant regions and labelled $U_i$ so that 
$U_{13}$ corresponds to the set $x_2=0$, $U_{14}$ the set $x_1=0$,  
$U_{15}$ the set $x_1+2x_3=0$ and $U_{16}$ the set $x_1=x_3$.   
In each of the remaining 12 regions of space a  
monotonic function has been identified.  These regions, and 
their corresponding monotonic functions, are given in Table \ref{5tab52}. 
Note that the totality of the sets $U_i$, $i=1..16$ provides a decomposition of the
 complete 
phase space.  As such, since each region is invariant under the system  
(\ref{5sys3a})-({\ref{5sys3b}), the existence of strictly monotonic functions in the 
regions $U_1$ - $U_{14}$ 
ensures that the only possible asymptotic solutions are located on the boundaries 
(either finite or infinite).  The finite boundaries are the sets 
$U_{13}-U_{16}$, or subsets thereof.  As a result the global dynamics has been completely  
determined by the previous investigations.  The only possible asymptotic states are, 
therefore, the singular points located at finite and infinite values. 
Furthermore, in the full four dimensional space the only possible asymptotic
states are those singular points which are sinks or sources; namely the sinks $A_+$, $B_-$ and $D_+$ 
and the sources $A_-$, $B_+$, and $D_-$. 

\subsection{The case $x_4=\pm \infty$}

>From table \ref{4tab42} we see that the only 
solutions characterized by $x_4\rightarrow \pm \infty$ correspond to the 
points $C_\pm$, found by compactifying the 
phase space using a Poincare transformation.  This point is 'non-hyperbolic' in
{\bf all} four directions.  To determine the exact nature of the local behavior 
of these points we consider 
the system  (\ref{5sys1a})-(\ref{5sys1b}) under the following change of coordinates:
\begin{equation}
Y_1=\frac{x_1}{x_4}, \quad Y_2=\frac{x_2}{x_4}, \quad Y_3=\frac{x_3}{x_4}, \quad 
\mbox{and} \quad Y_4=\frac{1}{x_4},
\end{equation}
and a "time" variable that is defined by $f'=Y_4\dot{f}$.  
The singular points of interest (namely  $C_\pm$) are now located at the origin
of the new coordinate system.  
In these new coordinates, the system
(\ref{5sys1a})-(\ref{5sys1b}) becomes:
\begin{eqnarray}
Y_1' &=& Y_1(Y_3-Y_2) \label{4sys4a} \\
Y_2' &=& Y_2(Y_4+Y_3-3Y_1-3Y_2) \\
Y_3' &=& Y_4+Y_1^2+Y_1Y_3-Y_2Y_3-Y_3^2 \\
Y_4' &=& -2Y_2Y_4. \label{4sys4b}
\end{eqnarray}
The singular points of interest (namely $C_\pm$) are now located at $(0,0,0,0)$.
There are two invariant lines, namely $Y_1=Y_3=Y_4=0$ and $Y_1=Y_2=Y_4=0$.
Each of these lines corresponds to an eigenvector of the flow for the
system (\ref{4sys4a})-(\ref{4sys4b}), and on each of these lines the flow is
monotonic decreasing. The dynamics in
a third direction can then be determined by considering the 
two-dimensional set
$Y_2=Y_4=0$, i.e.:
\begin{eqnarray}
Y_1' &=& Y_1Y_3 \label{4sys5a} \\
Y_3' &=& Y_1^2-Y_3^2+Y_1Y_3. \label{4sys5b}
\end{eqnarray}
In this case $Y_1=Y_3$ and $Y_1=-2Y_3$ are invariant sets (and, again, eigenvectors of the flow).  
On each of these
sets the derivatives are strictly positive or strictly negative, indicating
that this point is a saddle-node.

Therefore in three of the four directions (of the full phase space) 
through the singular points $C_\pm$
the derivative does not change sign, and in the fourth direction there
is no motion (as this direction is normal to the sheets of invariant planes
described in the previous section).  This point is, therefore, a higher-dimensional
saddle-node. 
 
\section{Description of asymptotic solutions} 
\setcounter{equation}{0}

In the qualitative analysis of the previous section, the asymptotic states of the 
governing system were described as singular points of the autonomous system of
ODEs.  The existence of other types of stable structures was ruled out by the 
existence of monotonic functions. 
While some of the singular points being described 
are not structurally stable in all 4-dimensions, there are invariant regions 
in which they do act as attractors (either to the past or the future);  
in addition, these points act as intermediate attractors (repellors) for 
large classes of solutions. 
Each of the physical solutions described by these 
singular points will now be given, restricting attention to those solutions 
which satisfy the weak and dominant energy conditions.   
In the case of infinite kinematic self-similarity, two of
the boundaries of the regions which satisfy the energy conditions 
are, in fact, invariant sets; therefore we need only consider the solutions 
which lie in the regions $x_1\geq 0$ and $x_4 + 2x_1x_3 + x_1^2 \leq 0$.
For the invariant set $x_1=0$, one has $\mu+p=0$ corresponding 
to a cosmological constant or vacuum if, in addition, $x_4=0$.
 
\subsection{Finite singular point asymptotic states} 

$Q_1=(0,1,0,0)$  is a vacuum solution corresponding to Minkowski 
space-time. $Q_2=(1,-1,1,0)$ does not satisfy the energy conditions since
$\mu-p<0$. $L_1=(0,0,\beta,\beta^2)$, if $\beta\neq 0$, then $\mu+p=0$ and
$\mu<0$ violating again the energy conditions. 
The remaining singular points are:
\begin{itemize}

\item $Q_3=(1/4,1/8,-1/8,0)$

In this case the metric is plane symmetric
\begin{equation} 
ds^2=-\left( {r\over t}\right)^{1/4}dt^2+
{b^2\over r^2}\left( {t\over r}\right)^{3/4}dr^2+ 
\left( {t\over r}\right)^{1/2}(d\theta^2+\theta^2d\phi^2) \ , \label{5metQ3} 
\end{equation} 
where $b$ is a constant. The energy density and pressure are given by
\be
\mu=p= {1\over 4 t^2} \left( {t\over r}\right)^{1/4} \ .
\ee
The fluid is physically self-similar and it represents the only
stiff-matter solution (i.e., $\mu=p$) in the plane symmetric case.

\item $L_2=(\sigma,0,0,-\sigma^2)$

The case $\sigma=0$ (i.e., the intersection of $L_1$ and $L_2$)
corresponds to Minkowski space-time. Otherwise, the metric is
 spherically symmetric
 \be
 ds^2=-dt^2 +{s_0^2\sigma^2\over r^2}\left( {t\over r}\right)^{2\sigma}
 dr^2 + s_0^2\left( {t\over r}\right)^{2\sigma}
 (d\theta^2+sin^2\theta d\phi^2) \ , \label{5metL2} 
 \ee
where $s_0$ and $\sigma$ are constants. The energy density and pressure are
\be
\mu={3\sigma^2 \over t^2}\ ,  \quad p={-3\sigma^2 +2 \sigma\over t^2}\ .
\ee  
In this case the fluid is physically self-similar and satisfies the
energy conditions  for $\sigma\geq 1/3$. The case $\sigma=1/3$ corresponds
to stiff-matter and $\sigma=2/3$ to dust.
\end{itemize}

\subsection{Infinite singular point asymptotic states} 
The infinite singular points are displayed in Table \ref{4tab42}. 
$A_\pm$, $B_\pm$ and $F_\pm$ correspond to vacuum solutions.
They asymptote respectively to the following line elements:
 \be
 ds^2(A)=-dt^2 
 +{b^2\over r^2} \left[ \ln \left({t\over r}\right) + c \right]^2 dr^2
 +d\theta^2+\theta^2d\phi^2 \ ,\quad {\rm for}\quad
 \ln \left({t\over r}\right) + c \rightarrow 0\ ,
 \ee
\be
 ds^2(B)=-\left[ \ln \left({t\over r}\right) + c \right]^2 dt^2 
 +{b^2\over r^2} dr^2 +d\theta^2+\theta^2d\phi^2 \ ,
 \quad {\rm for}\quad
 \ln \left({t\over r}\right) + c \rightarrow 0\ ,
\ee
\be
ds^2(F)=- \exp\left[c \left({t\over r}\right)^{1/2} \right] dt^2
+{b^2 t\over r^3} \exp\left[c \left({t\over r}\right)^{1/2} \right]dr^2
+d\theta^2+\theta^2d\phi^2 \ ,\quad {\rm for}\quad
{t\over r}\rightarrow \infty\ ,
\ee
 where $b$ and $c$ are integration constants.

The singular points $C_+$, $D_\pm$, $E_+$, $G_+$, and $H_\pm$ do not
satisfy energy conditions. Physical solutions (i.e., perfect
fluid solutions satisfying the energy conditions) do not asymptote
to
these  infinite singular points since they lie  in different invariant
regions of the phase space. The remaining  singular points are:
\begin{itemize}

\item $C_-$: The metric is given by
\be
ds^2=-dt^2 +{(\dot S)^2\over r^2}dr^2 + S^2 (d\theta^2+sin^2\theta d\phi^2) \ ,
\ee
where
\be
S=s_0\left({t\over r}\right)^{1/3}\left[\ln \left({t\over r}\right) + c
\right]^{2/3} \ ,\quad {\rm for}\quad
 \ln \left({t\over r}\right) + c \rightarrow 0\ ,
\ee
$s_0$ and $c$ are constants. The matter quantities are
\be
\mu={1\over 3t^2} \left[ 1 +{4 \over \ln \left( t/r\right) + c}\right] \ ,
\qquad p={1\over 3t^2} \ .
\ee
The fluid in this case is physically self-similar.

\item $E_-$: The metric is plane symmetric
\be
ds^2=-{1\over S} dt^2 +{b^2\over r^2}S^2 dr^2 
+S^2(d\theta^2+\theta^2d\phi^2) \ ,
\ee
where $S=[\ln(t/r)+c]^2$, for $\ln(t/r)+c\rightarrow 0$, $b$ and $c$
are constants. The matter variables are given by
\be
\mu={12\over t^2}\ , \qquad 
\mu+p={4\over t^2}\left[\ln\left({t\over r}\right)+c\right] \ .
\ee
Again the fluid is physically self-similar.

\item $G_-$: In this final case the metric is plane symmetric given by:
\be
ds^2=-{1\over S} dt^2 + {b^2\over r^2 S} dr^2 
+S^2(d\theta^2+\theta^2d\phi^2) \ ,
\ee
where $S=[\ln(t/r)+c]^{1/2}$, for $\ln(t/r)+c\rightarrow 0$,
$b$ and $c$
are constants.

Notice that this line element is not a solution of the system 
(\ref{5sys1a})-(\ref{5sys1b}). It is just the asymptotic solution when 
$\ln(t/r)+c\rightarrow 0$. For this reason we do not write the matter 
variables. All perfect fluid solutions approaching this singular point
will tend to be  physical self-similar. They will satisfy  the
energy conditions depending on the direction from which they are
approaching this point. $G_-$ lies exactly in the  boundary of a region
in which  the energy conditions are satisfied.
 \end{itemize}
In all the solutions here presented, $t$, $r$ and $\theta$  have been rescaled
in order to absorb as many integrating constants as possible.

\section{Special Cases}
\setcounter{equation}{0}
Having completed the qualitative analysis and identified the possible asymptotic 
states, it is useful to note that in several of the invariant sets considered
in the previous sections the system can be integrated completely so that the 
solutions can be written out explicitly.  These particular sets are considered
here.
 \subsection{The invariant set $x_2=0$}
 \label{sec6}
All of the exact solutions in this case can, in fact, be determined as
the system (\ref{5sys2a})-(\ref{5sys2b}) can be integrated completely.
If $x_1=0$, the remaining equation yields $\dot x_3=w_0 -x_3^2$.
But in this case  $\mu+p=0$ and perfect fluid solutions are excluded.
For $x_1\neq 0$ we have that
\be
x_3={\dot x_1\over x_1}  \ , 
\label{isx20}
\ee
and equation (\ref{5sys2b}) becomes
\be
\ddot x_1=x_1(w_0 + x_1^2 + \dot x_1) \   ,
\ee
that can be rewritten as
\be
(2\dot x_1 + x_1^2)\dot{}=2 x_1 (2\dot x_1 + x_1^2+w_0 ) \ ,
\label{isx21}
\ee
or
\be
(\dot x_1 -x_1^2)\dot{}= x_1 (-\dot x_1 + x_1^2+w_0 ) \ .
\label{isx22}
\ee
The case $\dot x_1=0$ and $x_1\neq 0$ implies $w_0 +x_1^2=0$ and
corresponds to the fixed points $L_2$. Apart from this case the
following possibilities arise.

\noindent
{\bf Case: $2\dot x_1 + x_1^2+w_0=0$.}

\noindent
Notice that all the solutions correspond to the intersection
of the invariants sets $I_2$ and $I_4$. The different solutions
depend on the value of $w_0$. They are:

\begin{itemize}

\item $w_0=0$: $x_1=2/\chi$, where $\chi\equiv \ln(t/r) +c$.
The metric can be written as
\be
ds^2=-{1\over \chi^2}dt^2 +{b^2 \chi^4\over r^2} dr^2 +\chi^4
(d\theta^2+\theta^2d\phi^2)  \ ,
\ee
where $b$ and $c$ are constants.

\item $w_0=+\beta^2$: $x_1=-\beta\tan(\beta\chi/2)$ and
\be
ds^2=-[\tan(\beta\chi/2)]^2dt^2+s_0^2[\cos(\beta\chi/2)]^4
\left( {\beta^2\over r^2} dr^2 + d\theta^2+ \sinh^2\theta d\phi^2\right) \ ,
\ee
$s_0$ being a constant.

\item $w_0=-\beta^2$: Two possibilities arise, 
$x_1=\beta\tanh(\beta\chi/2)$ or $x_1=\beta\coth(\beta\chi/2)$.
The line elements are
\be
ds^2=-[\tanh(\beta\chi/2)]^2dt^2+s_0^2[\cosh(\beta\chi/2)]^4
\left( {\beta^2\over r^2} dr^2 + d\theta^2+ \sin^2\theta d\phi^2\right) \ ,
\ee
and
\be
ds^2=-[\coth(\beta\chi/2)]^2dt^2+s_0^2[\sinh(\beta\chi/2)]^4
\left( {\beta^2\over r^2} dr^2 + d\theta^2+ \sin^2\theta d\phi^2\right) \ ,
\ee
respectively.
\end{itemize}
All of these solutions satisfy the energy conditions only over some limited regions of
the manifold.

\noindent
{\bf Case: $-\dot x_1 + x_1^2+w_0=0$.}

\noindent
Different solutions appear again depending on $w_0$. They correspond
to the intersection of the invariant sets $I_2$ and $I_5$.
\begin{itemize}

\item $w_0=0$: $x_1=-1/\chi$, and the metric can be written as
\be
ds^2={1\over \chi^2} \left(-dt^2 +{b^2\over r^2} dr^2 + 
d\theta^2+\theta^2d\phi^2 \right) \ .
\ee
The metric is conformally flat.

\item $w_0=+\beta^2$: $x_1=\beta\tan(\beta\chi)$,
\be
ds^2=-[\tan(\beta\chi)]^2dt^2 + {s_0^2\over [\cos(\beta\chi)]^2}
\left( {\beta^2\over r^2} dr^2 + d\theta^2+ \sinh^2\theta d\phi^2\right) \ .
\ee

\item $w_0=-\beta^2$: $x_1=-\beta\tanh(\beta\chi)$  or 
$x_1=-\beta\coth(\beta\chi)$.
\be
ds^2=-[\tanh(\beta\chi)]^2dt^2 + {s_0^2\over [\cosh(\beta\chi)]^2}
\left( {\beta^2\over r^2} dr^2 + d\theta^2+ \sin^2\theta d\phi^2\right) \ ,
\ee
and
\be
ds^2=-[\coth(\beta\chi)]^2dt^2 + {s_0^2\over [\sinh(\beta\chi)]^2}
\left( {\beta^2\over r^2} dr^2 + d\theta^2+ \sin^2\theta d\phi^2\right) \ ,
\ee
respectively.
\end{itemize}
In all cases $b$, $\beta$ and $s_0$ are constants.

\noindent
{\bf Case: $2\dot x_1 + x_1^2+w_0\neq 0$  and $-\dot x_1 + x_1^2+w_0\neq 0$.}

\noindent
Equations (\ref{isx21}) and  (\ref{isx22}) can be integrated to yield
\be
2\dot x_1 + x_1^2+w_0= 3\sigma S^2 \ ,
\label{isx23}
\ee
and
\be
-\dot x_1 + x_1^2+w_0={3 \lambda \over 2}{1\over S} \ ,
\label{isx24}
\ee
where $\sigma$ and $\lambda$ are arbitrary non-null constants.
Comparing  equations (\ref{isx23}) and (\ref{isx24}) we find
\be
x_1^2=\sigma S^2 +{\lambda \over S} -w_0 \ .
\label{isx25}
\ee
Hence, once $S$ is know, all the other metric functions can be calculated.
>From (\ref{isx20}) we get
\be
e^{2\Phi}=x_1^2 \ ,
\ee
and also we have
\be
e^{2\Psi}=b^2 S^2 \ ,
\ee
where $b^2=-w_0 k$ for $k\neq 0$ and any arbitrary constant for $k=0$.
Then substituting  $x_1=\dot S/S$ in equation (\ref{isx25}) one gets
\be
{\dot S\over \sqrt{\sigma S^4 -w_0 S^2 +\lambda S}}=\pm 1 \ .
\ee
This is the only equation that needs to be integrated. Thus, this case
$x_2=0$ is completely solved up to quadratures.

\subsection{The case $x_3=0$ (and $\dot x_3 =0$)}
 \label{sec7}
This case is of particular interest since all the solutions belong
to the intersection of the invariant sets $I_4$ and $I_5$ and
therefore they are physically self-similar. 
They represent the geodesic solutions for the system; i.e., these solutions
have zero acceleration. Since the governing equations impose the condition
$x_4=-x_1^2$, there can be no hyperbolically symmetric solutions in 
this case.  The plane symmetric case is the special case $x_1=0$
and this corresponds to vacuum solutions. Their metric is given by
\be
ds^2=-dt^2 +\left(1 +a {t\over r}\right)^2 {b^2\over r^2}dr^2 +
d\theta^2+\theta^2d\phi^2 \ ,
\ee
where $a$ and $b$ are constants.

All the other solution will exhibit spherical symmetry, and the
metric can be written as
\begin{equation} 
ds^2=-dt^2+\frac{\dot{S}^2}{r^2}dr^2+S^2(d\theta^2+\sin^2 
\theta d\phi^2) \ .
\end{equation} 
The governing equations reduce to 
\bea
\dot x_1&=& x_1 x_2 \ , \label{7s1}\\
\dot x_2& = &x_2(1-3x_1-x_2) \ , \label{7s2}
\eea
and substituting $x_2$ from (\ref{7s1}) into equation (\ref{7s2}),
we obtain
\be
\ddot x_1= -3x_1\dot x_1 + \dot x_1 \ .
\ee
A first integral this equation is
\be
\dot x_1=-{3\over 2} \left[\left(x_1-{1\over 3} \right)^2+n \right] \ ,
\ee
where $n$ is an arbitrary constant. The special case $\dot x_1=0$ 
corresponds to the curve $L_2$ of singular points. The other solutions
depend on the different value of the constant $n$ and they are
\begin{itemize}
\item $n=0$: $x_1=1/3+1/\chi$; $S=s_0 (t/r)^{1/3}\chi^{2/3}$, 
where now $\chi\equiv (3/2)(\ln(t/r)+c)$,
$c$ being a constant.

\item $n=\beta^2$: $x_1= 1/3 -\beta\tan(\beta\chi)$; 
$S=s_0 (t/r)^{1/3}[\cos(\beta\chi)]^{2/3}$.

\item $n=-\beta^2$: $x_1= 1/3 +\beta\tanh(\beta\chi)$; 
$S=s_0 (t/r)^{1/3}[\cosh(\beta\chi)]^{2/3}$,

 or
$x_1= 1/3 +\beta\coth(\beta\chi)$; 
$S=s_0 (t/r)^{1/3}[\sinh(\beta\chi)]^{2/3}$.
\end{itemize}
The solutions satisfy the energy conditions if $x_1 >0$ and
$x_1\geq 1/3 +3n$.



\section{Physical self-similarity} 
\setcounter{equation}{0} 
 As was stated in the introduction, the cases of physical and
geometric self-similarity are not necessarily equivalent
when considering KSS spacetimes.
Perfect fluid solutions will necessarily be physically self-similar
 if they satisfy 
 $ x_4+2x_1x_3+x_1^2=0$, and hence lie in the invariant set $I_4$.
 In this case the system (\ref{5sys1a})-(\ref{5sys1b}) 
 reduces to the  
three-dimensional system of autonomous ODEs: 
\begin{eqnarray} 
\dot{x_1} &=& x_1(x_2+x_3) \ , \label{5sys40a}  \\ 
\dot{x_2} &=& x_2(1-3x_1-x_2+x_3)\ ,  \\ 
\dot{x_3} &=& x_3(-x_1+x_2-x_3)  \ . \label{5sys40b} 
\end{eqnarray} 
Through the use of monotonic functions it 
can be shown that all of the asymptotic behavior in this class 
of solutions is described by solutions in one (or more) of 
the invariant sets $x_1=0$, $x_2=0$, $x_3=0$ or $x_1+2x_3=0$, 
all of which have been previously discussed. 
The case $x_1=0$ corresponds to the vacuum case, and the exact solutions for the
cases $x_2=0$ and $x_3=0$ can be found in section \ref{sec6} and \ref{sec7},
respectively.

In the case of physical self-similarity, perfect fluid solutions have a barotropic equation of
state if, in addition, they satisfy
\be
2x_2 + 3x_1= c_1 \ ,
\ee
where $c_1$ is an arbitrary constant. In Table \ref{8tab1}, 
we summarize all the
possible barotropic, physical self-similar solutions.

\section{Discussion} 
 
 To summarize, we have studied perfect fluid (spherically, plane
 and hyperbolically symmetric) space-times admitting a kinematic
 self-similarity of infinite type. 
 We have restricted our attention to the case in which the kinematic
 self-similar vector field commutes
 with all of the the Killing vectors.
 Three different cases arise depending
 on the orientation  of the fluid flow relative 
 to the kinematic self-similar vector.
The interesting general case is the \lq tilted' one in which  the four-velocity is
neither parallel nor orthogonal  to the self-similar vector field. 
In this case, we have shown that the governing  
equations reduce to a four-dimensional
autonomous system of ODEs.
The qualitative properties of the system have been fully studied.
In particular, through an extensive use of monotonic functions, we have shown
that all asymptotic solutions in this infinite class of 
kinematic self-similarity
are necessarily located at singular points (either at finite 
or infinite values of the dependent variables) which are classified 
in Tables \ref{5tab51} and \ref{4tab42}.  

Most of these singular points are saddle points
in the
full phase space, although there are invariant regions  in which they do act
as sinks or sources, thereby acting as attractors (or repellors)
for classes of solutions.  The {\bf only} global sinks and sources
are located on the infinite boundary, summarized in Table \ref{summ}.  Hence, 
in general solutions asymptote to one of those represented by the points
{\bf A}$_+$, {\bf B}$_+$ or {\bf D}$_-$ in the past and one of those
represented by the points {\bf A}$_-$, {\bf B}$_-$ or {\bf D}$_+$ in the future.

The physical solutions described by these singular points are given 
in the cases in which the weak and dominant energy conditions are satisfied.
The class of solutions which are also {\it physically} self-similar 
are again important in this analysis.  We show that 
in all cases in which the energy conditions are satisfied the 
asymptotic behavior is necessarily physical self-similar and
the space-time is plane or spherically symmetric.
This result coincides with the results of Benoit and Coley \cite{Benoit96},
which studied the case of spherical symmetry with finite 
kinematic self-similarity.  This again shows the relevance  of the 
physical self-similar  models.

In some special cases, e.g., the invariant set $x_2=0$, and the
geodesic case, corresponding to the case $x_3=0$, the four-dimensional
autonomous system of ODEs can be integrated completely. 
All the exact solutions  have been found in these cases (see Sections
6 and 7). In the geodesic case, the solutions are again physically
self-similar. These exact solutions serve as illustrations of the more general
qualitative results previously discussed.
Finally, we have also found all of the  physical self-similar solutions that
 admit a barotropic equation of state. The results are summarized
in Table 4.


\begin{center} 
\begin{table}\caption{\label{5tab51}  
   \textbf{Finite Singular Points} 
   for equations (\ref{5sys1a})-(\ref{5sys1b}).  The local  
analysis for each singular point will be discussed in the subsequent 
sections, according to their classification.  ${\bf Q}_{1,2,3}$  
are isolated singular points and ${\bf L}_1$ and 
${\bf L}_2$ are curves of non-isolated singular points  
(one-dimension equilibrium sets).  Note that the curves 
${\bf L}_1$ and ${\bf L_2}$ intersect at the point $(0,0,0,0)$,
which is a saddle-node.   
 $\beta$ and $\sigma$ are constants.} 
\begin{tabular}{|cc|cc|c|} 
\hline 
 & $(x_1, x_2, x_3, x_4)$ &  
\multicolumn{2}{|c|}{Eigenvalue - Eigenvector Pairs} & Classification   
 \\ \hline \hline 
${\bf Q}_1$ & $(0,1,0,0)$ & $1$ & $(0,1,2,0)$ & Saddle  \\ 
        & & $1$ & $(1,0,3,0)$ &  \\   
        & & $-1$& $(0,1,0,0)$ &  \\ 
        & & $2$ & $(0,1,3,3)$ &   \\ \hline \hline 
${\bf Q}_2$ & $(1,-1,1,0)$ & $-3$& $(5,-6,-9,0)$ & Saddle  \\ 
        & & $-2$& $(-4,5,3,7)$ &  \\ 
        & & $1+\sqrt{2}$& $(1,\sqrt{2},1,0)$ &  \\ 
        & & $1-\sqrt{2}$& $(1,-\sqrt{2},1,0)$ &  \\ \hline \hline 
${\bf Q}_3$ & $(\frac{1}{4},\frac{1}{8},\frac{-1}{8},0)$ & $3/4$& $(1,0,3,0)$ & Saddle \\ 
        & & $1/4$& $(-2,1,-3,2)$ &  \\ 
        & & $-(1-\sqrt{7}i)/8$& $(-2,-\sqrt{7}i,1,0)$ &  \\ 
        & & $-(1+\sqrt{7}i)/8$& $(-2,\sqrt{7}i,1,0)$  &  \\ \hline \hline 
${\bf L}_1$ & $(0,0,\beta,\beta^2)$ & $\beta$& $(3,0,1,0)$ & Curve of Saddle Points  \\ 
        & & $1+\beta$& $(0,1+\beta,\beta,2\beta^2)$ &   \\ 
        & & $-2\beta$& $(0,0,1,0)$ &    \\ 
        & & $0$& $(0,0,1,2\beta)$ &  \\ \hline \hline 
${\bf L}_2$ & $(\sigma,0,0,-\sigma^2)$ & $-\sigma$& $(-1,0,1,0)$& Curve of Saddle Points \\ 
        & & $2\sigma$& $(1,0,2,0)$ &   \\ 
        & & $1-3\sigma$& $(\sigma,-2\sigma,0,-3\sigma)$ & \\ 
        & & $0$& $(1,0,0,-2\sigma)$  & \\ \hline \hline 
\end{tabular} 
\end{table} 
\end{center}

\newpage

\begin{center} 
\begin{table}\caption[\textbf{Classification of the Infinite Singular Points}]
{\label{4tab42} 
   \textbf{Classification of the Infinite Singular Points} 
    of the Poincare transformed 
system (\ref{5sys1a})-(\ref{5sys1b})].  The exact solutions for each of
 these points are described in section 5.2.}
\begin{tabular}{|cc|cc|l|} 
\hline 
 & $(X_1, X_2, X_3, X_4)$ & 
\multicolumn{2}{|c|}{Eigenvalue - Eigenvector Pairs} 
& Classification \\ \hline \hline 
{\bf A}$_\pm$ & $(0,\pm 1,0,0)$ 
      & $\pm 2$ & $(1,0,0,0)$ & \\ 
        && $\pm 2$ & $(0,1,0,0)$ & {\bf A}$_+$ : Source \\    
      & & $\pm 2$& $(0,0,1,0)$ & \\ 
&       & $\pm 3$ & $(0,0,0,1)$ & {\bf A}$_-$ : Sink \\ \hline \hline 
{\bf B}$_\pm$ & $(0,0,\pm 1,0)$ 
      & $\pm 2$ & $(1,0,0,0)$ & \\ 
&       & $\pm 2$ & $(0,1,0,0)$ & {\bf B}$_+$ : Source \\ 
      & & $\pm 2$& $(0,0,1,0)$ & \\ 
&       & $\pm 1$ & $(0,0,0,1)$ & {\bf B}$_-$ : Sink \\ \hline \hline 
{\bf C}$_\pm$ & $(0,0,0,\pm 1)$ 
      & $0$& $(1,0,0,0)$& \\ 
        &&$0$& $(0,1,0,0)$ & Saddle-Node \\ 
      & & $0$& $(0,0,1,0)$& \\ 
        && $0$& $(0,0,0,1)$ &  \\ \hline \hline 
{\bf D}$_\pm$ & $\frac{\sqrt{2}}{2}(\pm 1,0,\pm 1,0)$ 
      & $\mp \sqrt{2}$ & $(1,0,1,0)$ & \\ 
&       & $\mp \frac{\sqrt{2}}{2}$ & $(0,0,0,1)$& {\bf D}$_+$ : Sink \\ 
      & & $\mp \frac{3\sqrt{2}}{2}$& $(-1,0,1,0)$ & \\ 
&       & $\mp \frac{3\sqrt{2}}{2}$ & $(0,1,0,0)$ & {\bf D}$_-$ : Source \\ \hline \hline 
{\bf E}$_\pm$ & $\frac{\sqrt{5}}{5}(\mp 2,0,\pm 1,0)$ 
      & $\pm \frac{6\sqrt{5}}{5}$ & $(0,1,0,0)$ & \\ 
&       & $\mp \frac{6\sqrt{5}}{5}$ & $(1,0,2,0)$ & Saddle \\ 
      & & $\mp \frac{\sqrt{5}}{5}$& $(0,0,0,1)$ & \\ 
&       & $\mp \frac{2\sqrt{5}}{5}$ & $(-2,0,1,0)$ & \\ \hline \hline     
{\bf F}$_\pm$ & $\frac{\sqrt{2}}{2}(0,\pm 1,\pm 1,0)$ 
      & $\mp \sqrt{2}$& $(0,-1,1,0)$ & \\ 
&       & $\pm \sqrt{2}$& $(2,-1,1,0)$ & Saddle \\ 
      & & $\pm \sqrt{2}$& $(0,0,0,1)$ & \\ 
&       & $0$& $(0,1,1,0)$ & \\ \hline \hline 
{\bf G}$_\pm$ & $\frac{\sqrt{14}}{14}(\mp 2,\pm 3,\pm 1,0)$ 
      & $\pm \frac{\sqrt{14}}{7}$& $(0,0,0,1)$ & \\ 
&       & $\mp \frac{4\sqrt{14}}{7}$& $(2,-3,-1,0)$ & Saddle \\ 
      & & $\mp \frac{3\sqrt{14}}{7}$& $(1,0,2,0)$ & \\ 
&       & $\mp \frac{3\sqrt{14}}{7}$& $(0,1,-3,0)$ & \\ \hline \hline 
{\bf H}$_\pm$ & $\frac{\sqrt{17}}{17}(\pm 1,\mp 3/2,\pm 1,0)$ 
      & $\mp \frac{27\sqrt{17}}{136}$& $(134,-201,257,0)$ & \\ 
&       & $\mp \frac{\sqrt{17}}{34}$& $(2,-3,2,0)$ & Saddle \\
      & & $\pm \frac{21\sqrt{17}}{136}$& $(63,118,63,0)$& \\
&       & $\mp \frac{23\sqrt{17}}{136}$& $(0,0,0,1)$ & \\ \hline \hline 
\end{tabular} 
\end{table} 
\end{center} 

\newpage

\begin{center} 
\begin{table}\caption{\label{5tab52} 
        \textbf{Invariant regions of the space} $(x_1,x_2,x_3)$ for the system
         (\ref{5sys3a})-(\ref{5sys3b})    
        with corresponding monotonic functions.  Note that the 
        sets $U_{13}-U_{16}$ are not included here as they are two-dimensional
        invariant sets and their complete dynamics 
        have been summarized in the phase portraits: 
        Figures \ref{fig:45a}-\ref{fig:54} .} 
\begin{tabular}{|l|c|c|} 
\hline s
Label & Definition of Region & Monotonic \\ 
        &               &       Function \\ \hline \hline 
$U_1$ & $\{(x_1,x_2,x_3)|x_1,x_2,x_3 >0 \} $ & $x_1$ strictly increasing \\ \hline 
$U_2$ & $\{(x_1,x_2,x_3)|x_1,x_2 >0, x_1+2x_3>0 \} $ & $x_1+2x_3$ strictly increasing \\ \hline 
$U_3$ & $\{(x_1,x_2,x_3)|x_1,x_2 >0, x_1+2x_3<0 \} $ & $x_1+2x_3$ strictly decreasing \\ \hline 
$U_4$ & $\{(x_1,x_2,x_3)|x_1<0,x_2 >0, x_1-x_3<0 \} $ & $x_1-x_3$ strictly decreasing \\ \hline 
$U_3$ & $\{(x_1,x_2,x_3)|x_1<0,x_2 >0, x_1-x_3>0 \} $ & $x_1-x_3$ strictly increasing \\ \hline 
$U_6$ & $\{(x_1,x_2,x_3)|x_1<0, x_2,x_3>0 \} $ & $x_1$ strictly decreasing \\ \hline 
$U_7$ & $\{(x_1,x_2,x_3)|x_1>0, x_2<0, x_1-x_3<0 \} $ & $x_1-x_3$ strictly increasing \\ \hline 
$U_8$ & $\{(x_1,x_2,x_3)|x_1>0,x_2 <0, x_1-x_3>0 \} $ & $x_1-x_3$ strictly decreasing \\ \hline 
$U_9$ & $\{(x_1,x_2,x_3)|x_1>0, x_2<0,x_3<0 \} $ & $x_1-x_3$ strictly decreasing \\ \hline 
$U_{10}$ & $\{(x_1,x_2,x_3)|x_1,x_2,x_3 <0 \} $ & $x_1$ strictly increasing \\ \hline 
$U_{11}$ & $\{(x_1,x_2,x_3)|x_1,x_2<0, x_1+2x_3 <0 \} $ & $x_1+2x_3$ strictly increasing \\ \hline 
$U_{12}$ & $\{(x_1,x_2,x_3)|x_1,x_2<0, x_1+2x_3>0 \} $ & $x_1+2x_3$ strictly decreasing \\ \hline \hline 
\end{tabular} 
\end{table} 
\end{center}

 \newpage
\begin{center} 
\begin{table}\caption{\label{8tab1} 
        \textbf{Infinite kinematic self-similar perfect fluid
        solutions admitting a barotropic equation of state.} 
        The weak and dominant energy conditions are always 
        satisfied,
         except in the  last case in which  $c_1\geq 1$.} 
\begin{tabular}{|l|c|l|} 
\hline 
Case & Solution & Remarks \\  \hline \hline 
$c_1=0$ & $x_1=0$, $x_2=0$, $x_3=[\ln(t/r) +c]^{-1}$, $x_4=0$
 & vacuum\\ \hline
$c_1=1$ & $x_1=1/4$, $x_2=1/8$, $x_3=-1/8$, $x_4=0$ & 
$Q_3$: stiff-matter\\ \hline
 $c_1=1$ & $x_1=1/3$, $x_2=0$, $x_3=0$, $x_4=-1/9$ & 
$L_2$: stiff-matter\\ \hline
 $c_1=2$ & $x_1=0$, $x_2=1$, $x_3=0$, $x_4=0$ & 
$Q_1$:  vacuum\\ \hline
 $c_1=2$ & $x_1=2/3$, $x_2=0$, $x_3=0$, $x_4=-4/9$ & 
$L_2$: dust\\ \hline
 $c_1=2$ & $x_3=0$, $x_4=-x_1^2$, $x_2=1-3 x_1/2$ & \\
          & see case $x_3=0$, with $\beta=1/3$, $n=-1/9$ & dust\\ \hline
 $c_1\neq 0,1,2$ & $x_1=c_1/3$, $x_2=0$, $x_3=0$, $x_4=-c_1^2/9$ & 
$L_2$\\ \hline
\end{tabular} 
\end{table} 
\end{center}

\newpage

\begin{center} 
\begin{table}\caption{\label{summ} 
        \textbf{Summary of the global asymptotic behaviour.} }
\begin{tabular}{|l|l|l|} 
\hline 
Singular Point & Nature of the Singular Point & Physical Characteristics of Solutions \\ \hline \hline
{\bf A}$_+$ & Source:  Global attractor to the past & Geodesic, vacuum solution; \\
	    &                                       &  physically self-similar \\ \hline
{\bf A}$_-$ & Sink: Global attractor to the future & Geodesic, vacuum solution; \\
            &                                      &  physically self-similar \\ \hline
{\bf B}$_+$ & Source: Global attractor to the past & Vacuum solution; \\
            &                                      & physically self-similar \\ \hline
{\bf B}$_-$ & Sink: Global attractor to the future & Vacuum solution; \\
            &                                      & physically self-similar \\ \hline
{\bf D}$_+$ & Sink: Global attractor to the future & Energy conditions not satisfied \\ \hline
{\bf D}$_-$ & Source: Global attractor to the past & Energy conditions not satisfied \\ \hline
\end{tabular} 
\end{table} 
\end{center}


\newpage

\begin{figure} 
 \centering 
 \epsfig{file=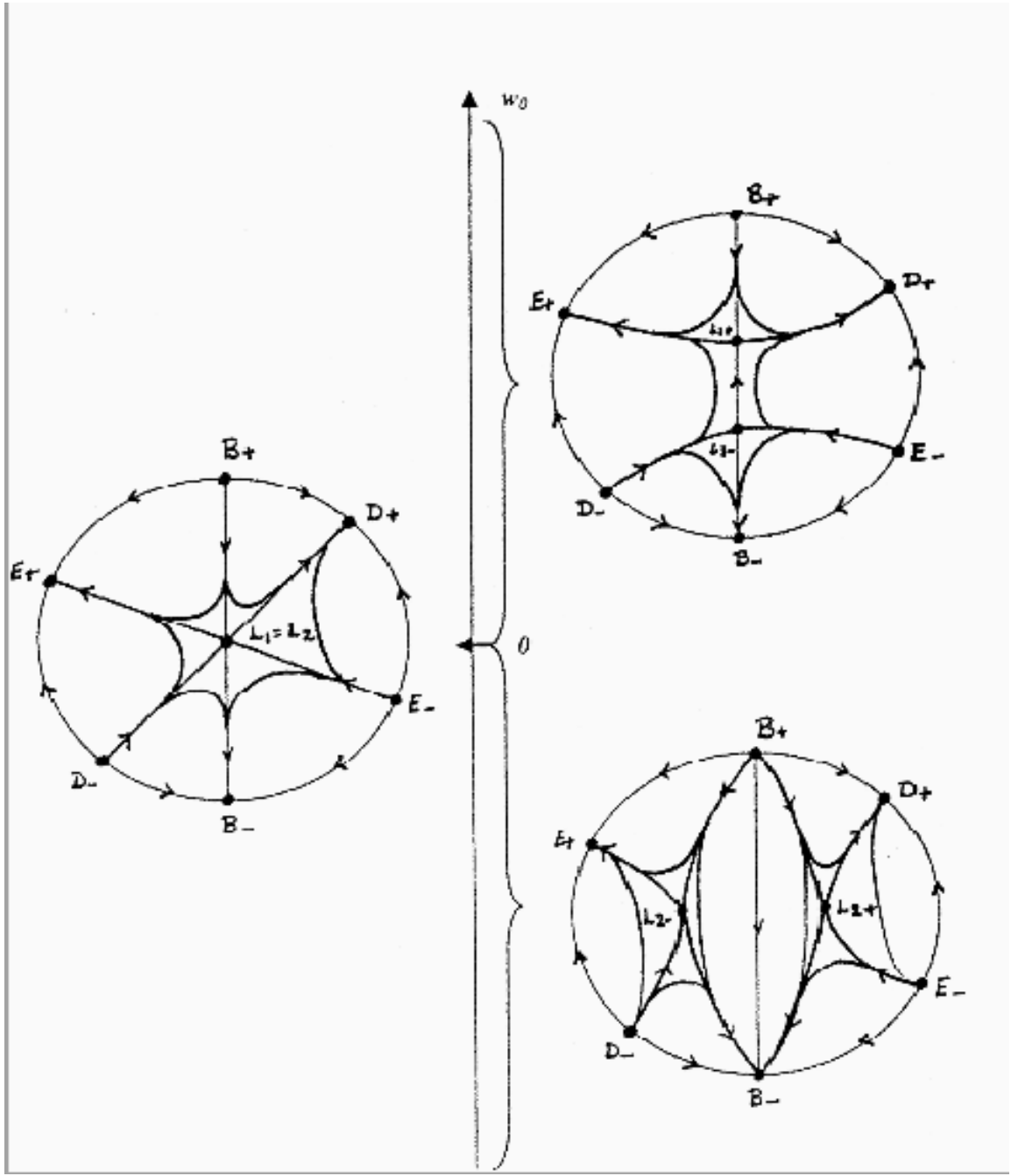,height=7in} 
 \caption[Phase portrait of the Poincare transformed system (\ref{5sys2a})- 
        (\ref{5sys2b})]{ 
     \label{fig:51} 
     Phase portrait of the Poincare transformation of system (\ref{5sys2a})- 
        (\ref{5sys2b}): 
        {\em The case $w_0<0$ corresponds to spherically symmetric solutions, 
        $w_0=0$ to plane symmetric solutions, and $w_0>0$ to hyperbolically symmetric 
        solutions.  The vertical direction represents the parameter space 
        $w_0 \in \Re$, where $w_0=0$ is the bifurcation value for the system.}
 } 
\end{figure}

\newpage

\begin{figure} 
 \centering 
 \epsfig{file=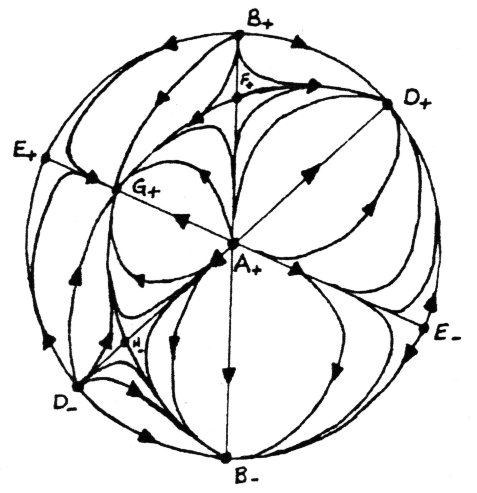,height=3.5in} 
 \caption[Dynamics on the infinite boundary for the plane symmetric solutions]
{ 
     \label{fig:45a} 
     Dynamics on the infinite boundary for the plane symmetric solutions: 
        {\em for the Poincare transformed space. Top hemisphere; i.e. $X_2>0$} 
 } 
\end{figure} 
 
\newpage

\begin{figure} 
 \centering       
 \epsfig{file=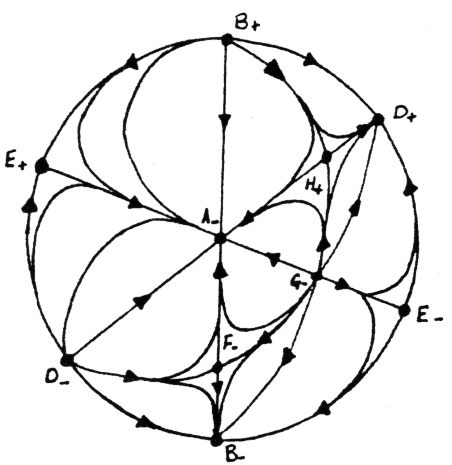,height=3.5in} 
 \caption[Dynamics on the infinite boundary for the plane symmetric solutions]
{ 
     \label{fig:45b} 
     Dynamics on the infinite boundary for the plane symmetric solutions: 
        {\em for the Poincare transformed system. Bottom hemisphere; i.e. $X_2<0$
} 
 } 
\end{figure}   

\newpage

\begin{figure} 
 \centering 
 \epsfig{file=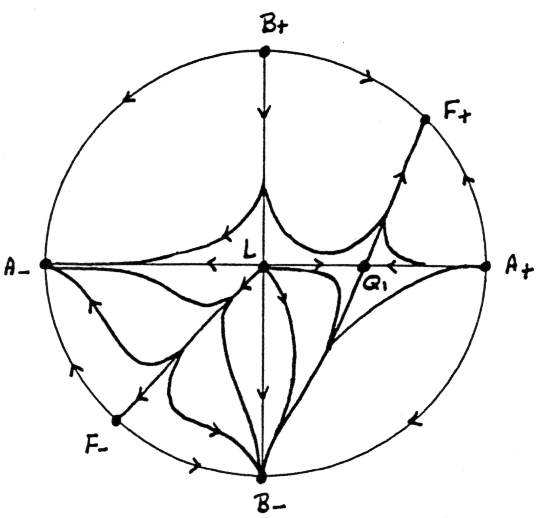,height=3.5in} 
 \caption[Phase portrait of the Poincare transformation of system (\ref{5x13a}
)- 
        (\ref{5x13b})]{ 
     \label{fig:52} 
     Phase portrait of the Poincare transformed system (\ref{5x1a})- 
        (\ref{5x1b}): 
        {\em plane symmetric solutions restricted to the invariant set  
        $x_1=0$.  The phase space is $X_2$ vs. $X_3$.} 
 } 
\end{figure} 
 
\newpage

\begin{figure} 
 \centering 
 \epsfig{file=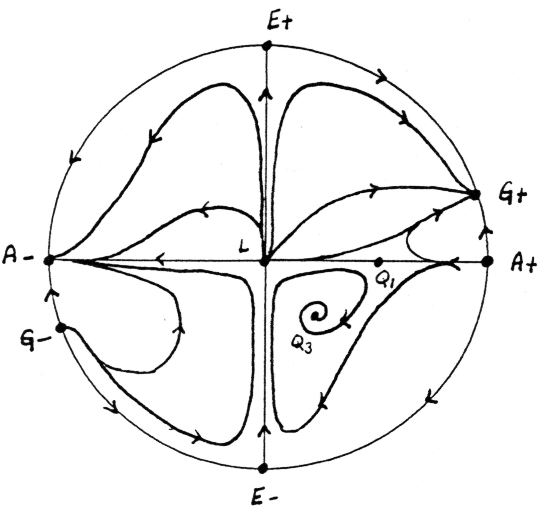,height=3.5in} 
 \caption[Phase portrait of the Poincare transformed system (\ref{5x13a}
)- 
        (\ref{5x13b})]{ 
     \label{fig:53} 
     Phase portrait of the Poincare transformed system (\ref{5x13a})- 
        (\ref{5x13b}): 
        {\em plane symmetric solutions restricted to the invariant set  
        $x_1+2x_3=0$.  The phase space is $X_2$ vs. $X_3$.} 
 } 
\end{figure}

\newpage

\begin{figure} 
 \centering 
 \epsfig{file=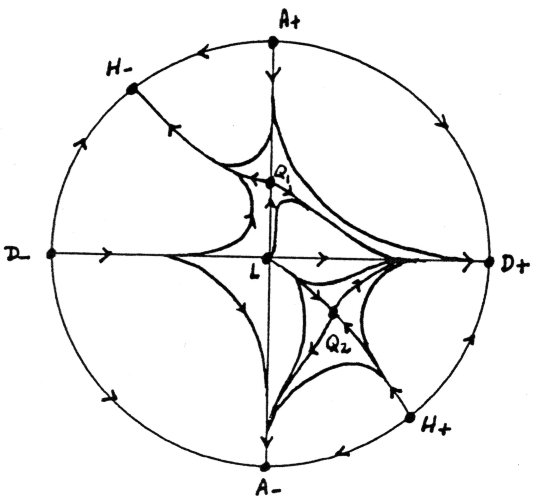,height=3.5in} 
 \caption[Phase portrait of the Poincare transformation of system
           (\ref{5x3a})-(\ref{5x3b})]{ 
     \label{fig:54} 
     Phase portrait of the Poincare transformed system (\ref{5x3a})- 
        (\ref{5x3b}): 
        {\em plane symmetric solutions restricted to the invariant set  
        $x_1=x_3$.  The phase space is $X_1$ vs. $X_2$.} 
 } 
\end{figure}


\begin{thebibliography}{99} 
\bibitem{Sedov} L.I. Sedov, {\em Similarity and Dimensional Methods in 
Mechanics}, Academic Press, New York (1959). 
\bibitem{Bare72} G.E. Barenblatt and Ya. B. Zeldovich, 1972, 
Ann. Rev. Fluid Mech., {\bf 4}, 285. 
\bibitem{Cahill71} M.E. Cahill and A H. Taub, 1971, 
 Comm. Math. Phys. {\bf 21}, 1. 
\bibitem{Eard1} D.M. Eardley, 1974, Comm. Math. Phys. {\bf 37}, 287. 
\bibitem{Eard2} D.M. Eardley, 1974, Phys. Rev. Lett. {\bf 33}, 442. 
\bibitem{Carter89}  B. Carter and R.N. Henriksen, 
1989, Annales de Physique, Paris Suppl. no 6, {\bf 14}, 47. 
\bibitem{Carter91}  B. Carter and R.N. Henriksen, 1991, J. Math. Phys.  
{\bf 32}, 2580. 
\bibitem{Coley96}  A. A. Coley, 1997, Class. Quant. Grav. {\bf 14}, 87.
\bibitem{Sintes98} A. M. Sintes, 1998, Class. Quant. Grav. {\bf 15}, 3689.

\bibitem{Benoit96} P. M. Benoit and A. A. Coley, 1998, Class. Quant. Grav. 
{\bf 15}, 2397.
\bibitem{Benoit98} P. M. Benoit, 1999, {\it PhD Thesis}, Dalhousie University 
\bibitem{Coley94} A. A. Coley and B. O. J. Tupper, 1994, 
 Class. Quant. Grav. {\bf 11}, 2553. 
\bibitem{CMT} J. Guckenheimer and P.  Holmes, 1983, Nonlinear Oscillations, 
 Dynamical Systems, and Bifurcations (Wiley) 
\bibitem{LeBlanc} V. G. LeBlanc, D. Kerr, and J. Wainwright, 1995, Class.  
 Quant. Grav, {\bf 12}, 513 
\bibitem{Bell}  D. Lynden-Bell and J. P. S. Lemos, 1988, 
 Mon. Not. R. Ast. Soc. {\bf 233}, 197. 
 
\bibitem{M}  R. Maartens, D. P. Mason and M. Tsamparlis, 1986, 
 J. Math. Phys. {\bf 27}, 2987. 
 
\end{thebibliography}
\end{document}